\documentclass[aps,preprint]{revtex4}

\usepackage{amssymb,amsmath,amsthm,graphicx}

\newtheorem{theorem}{Theorem}[section]
\newtheorem{lemma}[theorem]{Lemma}

\theoremstyle{definition}
\newtheorem{definition}[theorem]{Definition}
\newtheorem{example}[theorem]{Example}

\theoremstyle{remark}

\begin{document}

\title{Spreading of infectious diseases on heterogeneous populations:\\
multi-type network approach}

\author{Alexei Vazquez$^{1,2}$}

\affiliation{$^1$Center for Cancer Systems Biology, Dana Farber Cancer 
Institute, Harvard Medical School, 44 Binney St, Boston, MA 02115, USA\\
$^2$Department of Physics and Center for Complex Networks Research, 
University of Notre Dame}

\date{\today}

\begin{abstract}

I study the spreading of infectious diseases on heterogeneous populations.
I represent the population structure by a contact-graph where vertices
represent agents and edges represent disease transmission channels among
them. The population heterogeneity is taken into account by the agent's
subdivision in types and the mixing matrix among them. I introduce a
type-network representation for the mixing matrix allowing an intuitive
understanding of the mixing patterns and the analytical calculations.
Using an iterative approach I obtain recursive equations for the
probability distribution of the outbreak size as a function of time.  I
demonstrate that the expected outbreak size and its progression in time 
are
determined by the largest eigenvalue of the reproductive number matrix and
the characteristic distance between agents on the contact-graph. Finally,
I discuss the impact of intervention strategies to halt epidemic
outbreaks. This work provides both a qualitative understanding and tools
to obtain quantitative predictions for the spreading dynamics on
heterogeneous populations.

\end{abstract}

\maketitle

\bibliographystyle{apsrev}

\section{Introduction}

The globalization of human interactions have created a fertile ground for
the fast and broad spread of infectious diseases, potentially leading to
worldwide epidemics. We are thus force to understand the spreading of
infectious diseases within this global scenario. Yet, the study of
worldwide epidemics is challenging given the heterogeneity of the
populations involved 
\cite{anderson91,hufnagel04,koopman04,germann06,colizza06}.

The first sign of heterogeneity is given by the variability of the
reproductive number within or across populations
\cite{may87,anderson04,lloyd05}.  The reproductive number is defined as
the number of secondary cases generated by a primary infected case within
a population of susceptible individuals. In the case of sexually
transmitted diseases the reproductive number is proportional to the rate
of sexual partner acquisition \cite{may88,anderson91} and it exhibits wide
fluctuations \cite{may87,anderson91,liljeros01,jones03,schneeberger04}. In
network based approaches the reproductive number is proportional to the
node's degree \cite{pv00,mn00} and it exhibits wide fluctuations as well
\cite{ab01a}. In the absence of biases among the connections between
agents this heterogeneity is completely taken into account by the
reproductive number distribution \cite{pv00,mn00}.

There are other properties beyond the reproductive number requiring the
subdivision of a population in different classes or types. This includes
but is not limited to age, geographical location, social status and sexual
behavior. In general these heterogeneities cannot be characterized by a
single probability distribution. They require a multi-type approach with
probability distributions characterizing each type and a mixing matrix
describing the patterns of transmission among them.

Multi-type models are difficult to deal with and are generally tackled
using multi-agent simulations
\cite{rvachev85,flahault88,eubank04,hufnagel04,colizza06,germann06}. The
advantage of multi-agent simulations is that we can consider several
details and study their impact on the spreading dynamics. On the other
hand, given the large number of variables and model parameters it is
difficult to understand which are the key variables driving the system's
dynamics. Therefore, analytical calculations are required to funnel the
multi-agent simulations into specific regions of the parameters space.

In this work I study the spreading of infectious diseases on multi-type
networks. I take as starting point the static problem formulation
developed by Newman \cite{newman03} and the theory of age-dependent
multi-type branching process \cite{mode71}. I develop these mathematical
approaches to accommodate some distinctive properties of real networks
that have not previously considered. In section \ref{sec:multi} I
introduce the basic framework. Focusing on the population structure I
consider the contact-graph characterizing the detailed interactions among
agents and, at a metapopulation level, the type-network characterizing the
interactions among agent's types.  Through some simple examples I
illustrate the properties of the mixing matrix and its type-network
representation. This section ends defining a branching process modeling a
spanning tree from an index agent to all other agents in the
contact-graph. In section \ref{sec:spreading} I characterize the local
spreading dynamics from an agent to its contacts, taking the susceptible,
infected, and removed (SIR) model as a case study. Bringing together the
underlying network structure and the local transmission dynamics in
section \ref{sec:spreading} I define a branching process that models the
disease spreading dynamics. In section \ref{sec:iterative} I extend the
iterative approach for a single type
\cite{vazquez06a,vazquez06b,vazquez06e} to accommodate the particularities
of the multi-type case. Focusing on the expected behavior, in section
\ref{sec:average} I obtain general equations determining the progression
of the expected number of cumulative and new infections. Starting from
these equations I analyze some limited cases. First, I derive the final
expected outbreak size and, second, I analyze the time progression of the
expected outbreak size for the case of a time homogeneous local
transmission. In section \ref{sec:intervention} I discuss the impact of
the population heterogeneity on intervention strategies. I emphasize the
role of the characteristic distance between agents to quantify the impact
of intervention strategies on small-world populations.  I also illustrate
interventions targeting specific agent's types using a bipartite
population as a case study. Finally, in section \ref{sec:discussion} I
provide an overview of the main results and discuss future directions.

\section{Population structure}
\label{sec:multi}

Consider a population of $N$ agents that are susceptible to an infectious
disease. By {\it agent} I mean any entity that could host and transmit the
disease. Since we are interested on the transmission of infectious
diseases among humans an agent is a human in the first place.  For
vector-borne diseases we could have in addition agents representing the
intermediary host while for airborne diseases an agent could also
represent a public place. The agents are assumed to be heterogeneous
meaning that there are different agent classes or {\it types} according to
their pattern of connectivity to other agents and/or to the speed at which
they could potentially transmit an infectious disease. For instance, human
can be divided according to their age, social status and geographical
location.  Furthermore, in the case of vector- and air-born diseases there
is an additional type given by the non-human intermediary. More
precisely, let us assume that the agent population is divided in $M$
types and there are $N_a$ agents of type $a=1,\ldots,M$, satisfying the
normalization condition

\begin{equation}
\sum_{a=1}^M N_a = N\ .
\label{NiN}
\end{equation}

\noindent Note that within this work I use the indexes $a,b,\ldots$ for
the agent's type. In the following I introduce two representations of the 
population structure at the agent and type levels, respectively.

\subsection{Contact-graph}
\label{sec:cg}

The contact-graph takes precisely into account who could potentially
transmit the disease to whom
\cite{anderson91,friedman97,edmunds97,ghani98,keeling05}. More precisely,

\begin{definition}

The contact-graph is a labeled graph where vertices represent agents,
edges represent the potential disease transmission channels among them,
and the vertices are labeled according to the agent's type.

\label{cg}
\end{definition}

\noindent The contact-graph represents the population mixing at the
agent's level. Since there is a one-to-one relation between vertices and 
the corresponding agents I use these two terms interchangeable.

All the information necessary to characterize a given graph is provided by
its adjacency matrix. Yet, we should take into account the large size of
real populations and their change in time. In general, the only way to
achieve such a detailed description relies on agent-based simulations. My
scope is to bypass this detailed description and focus on statistical
properties that does not depend on the population structure details or
their change in time. Yet, to achieve that I need to specify the time 
scale where these statistical properties are measured. 

Excluding the effect of patient isolation or any other intervention, the
time scale that matters is the time interval from the infection of an
agent to its death or recovery, i.e. the disease life time within an
agent. At this point I intentionally exclude the effect of interventions,
such as patient isolation, in order to achieve a more general approach.
Their influence is taken into account when defining the disease spreading
dynamics (see section \ref{sec:spreading}). It is also worth mentioning
that the disease life time is a random variable. Therefore, the
statistical properties introduced below are the expectation after
averaging over the disease life time distribution.

The degree of a node is the total number of edges emanating from it
regardless the type of the node at the other end. Let $p^{(a)}_k$ be
probability distribution that a type $a$ node has degree $k$ and denote by

\begin{equation}
\langle k\rangle_a = \sum_{k=1}^\infty p_k^{(a)}k\ .
\label{meank}
\end{equation}

\noindent its mean. Note that by allowing $k$ to take values larger than
one we are already taking into account the existence of concurrency
\cite{watts92,kretzschmar96,garnett97}.

To characterize the spreading process it is also relevant to determine the
same distribution but for a vertex found and the end of an edge selected
at random. This sampling introduces a bias towards nodes with higher
degree resulting in the probability distribution

\begin{equation}
q^{(a)}_k = \frac{kp^{(a)}_k}{\sum_{s=1}^\infty sp^{(a)}_s}\ .
\label{qi}
\end{equation}

\noindent with average excess degree

\begin{equation}
\langle k\rangle_a^{\rm (excess)} = \sum_{k}q_k^{(a)}(k-1)\ .
\label{meanK}
\end{equation}

\noindent where the minus one subtracts the edge from where the node was
reached. Associated with these two probability distributions we introduce
the generating functions

\begin{equation}
U_a(x) = \sum_{k=0}^\infty p^{(a)}_k x^k\ ,
\label{Ui}
\end{equation}

\begin{equation}
V_a(x) = \sum_{k=1}^\infty q^{(a)}_k x^{k-1}\ .
\label{Vi}
\end{equation}

\noindent From the derivatives of $U_a(x)$ and $V_a(x)$ we obtain the
moments of $p_k^{(a)}$ and $q_k^{(a)}$, respectively. For instance

\begin{equation}
\dot{U}_a(1) = \langle k\rangle_a\ ,
\label{Umeank}
\end{equation}

\begin{equation}
\dot{V}_a(1) = \langle k\rangle^{\rm (excess)}_a\ .
\label{VmeanK}
\end{equation}

Since the agent population is finite there is a typical distance $D$
between every two agents on the contact-graph. Social experiments such as
the Kevin Bacon and Erd\H{o}s numbers \cite{w99} or the Milgram experiment
\cite{m67} reveal that social actors are separated by a small number of
acquaintances (``small-world'' property \cite{ws98}). This observation is
supported by theoretical approaches demonstrating that $D$ grows at most
as $\log N$ in random graphs
\cite{bollobas01,chung02,bollobas03,cohen03a}. More recently it has been
shown that for several real networks $D$ actually decreases or remains
constant as the network evolve and increases its size \cite{leskovec06}.  
Thus, I explicitly take into account that $D$ is finite.

\begin{example}[Poisson contact process]

Let us assume that type $a$ agents establish connections with other agents
at a constant rate $\lambda_a$ and that the disease life time is constant
and equal to $T$. In this case we obtain a Poisson distribution for the
agent's degree

\begin{equation}
p^{(a)}_k=\frac{(\lambda_a T)^ke^{-\lambda_a T}}{k!}\ .
\label{pkp}
\end{equation}

\noindent Furthermore, $q^{(a)}_k=p^{(a)}_k$, $\langle k\rangle_a=\langle 
k\rangle_a^{\rm (excess)}=\lambda_a T$, and $U(x)=V(x)=e^{(x-1)\lambda_a 
T}$.

\label{poisson1}
\end{example}

\begin{figure}
\centerline{\includegraphics[height=3in]{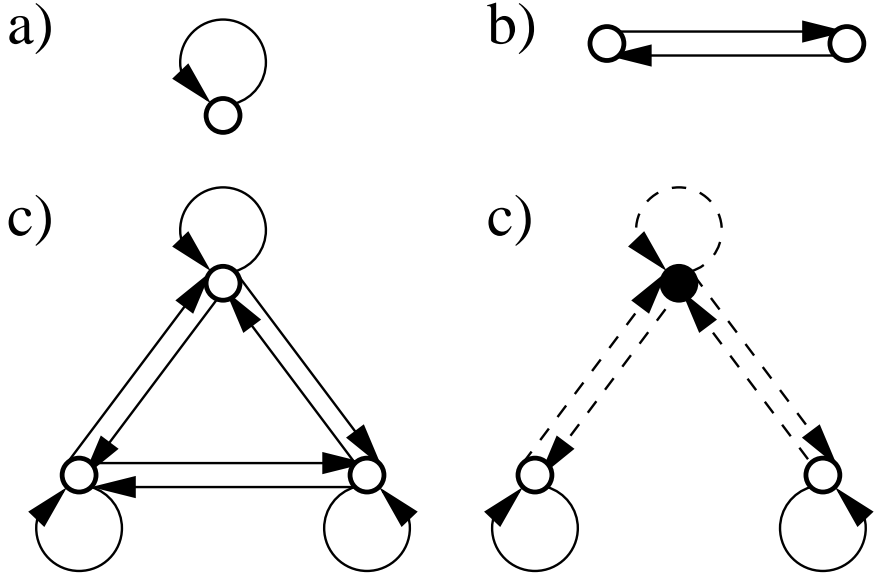}}

\caption{Type-network representation of simple mixing matrices. a)  
Single-type population. b) Bipartite population. c) Fully mixed population
with three types. c) Two cities (open circles) and the commuters among
them (solid circle). The continuous/dashed lines represent intra/inter
city connections.}

\label{typegraph}
\end{figure}

\subsection{Type-network}

At the metapopulation level the population structure of the is determined
by the mixing patterns among the different agent's types. Given a type $a$
agent and one of its edges let $e_{ab}$ be the probability that the agent
at the other end is of type $b$ ( {\it mixing matrix}). From the mixing
matrix we can construct the type-network characterizing the metapopulation
structure.

\begin{definition}

{\it Type-network:}  In the type-network a node represents a type,
an arc is drawn from type $a$ to $b$ if $e_{ab}>0$, and the arc's weights
are given by $e_{ab}$.  

\label{tn}
\end{definition}

\noindent Note that since $e_{aa}$ may be nonzero the type-network may
contain loops. Fig. \ref{typegraph} shows some simple type-networks. The
single-type case is represented by a node with a loop (Fig.
\ref{typegraph}a). A bipartite population is represented by two nodes with
an incoming and an outgoing arc (Fig. \ref{typegraph}b). This example
could model a heterosexual population with no other distinction than
gender or a metapopulation given by people and public places
\cite{eubank04}. A fully mixed population is represented by a complete
network (Fig. \ref{typegraph}c). A less intuitive example is the
type-network shown in Fig. \ref{typegraph}d, representing a population
divided in two cities and the commuters between them.

\subsection{Annealed spanning tree}
\label{sec:ast}

Given a contact-graph, let us consider an epidemic outbreak starting
from a single agent (index case). In the worth case scenario the disease
propagates to all the agents that could be reached from the index case
using the network connections. Thus, the outbreak is represented by a
spanning or causal tree from the index case to all reachable agents. On
this tree, the generation of an agent corresponds with the topological or
hopping distance from the index case. This picture motivates the
introduction of the following branching process:

\begin{definition}{Multi-type Annealed Spanning Tree (AST)}

Consider a labeled contact-graph characterized by $\{N_a,\ p^{(a)}_k\}$
and the type-network $\{e_{ab}\}$. The multi-type annealed spanning tree
(AST) is the branching process satisfying the following properties:

\begin{enumerate}

\item The process start from an index case of type $a\in\{1,\ldots,M\}$
at generation $d=0$. The index case generates $k$ sons
with probability distribution $p^{(a)}_k$.  Each son is of type $b$ with
probability $e_{ab}$.

\item Each son at generation $1\leq d<D$ generates $k-1$ sons with
probability distribution $q^{(a)}_k$. Each son is of type $b$ with
probability $e_{ab}$.

\item A son at generation $d=D$ does not generate new sons.

\end{enumerate}

\label{AST}
\end{definition}

\noindent The term annealed means that we are not analyzing the true
(quenched) spanning tree on the graph but a branching process with similar
statistical properties. This approximation is particularly good if the
contact-graph is continuously changing in time albeit the constancy of its
statistical properties. A similar mathematical construction has been
previously introduced by Newman \cite{n02a,newman03}. The main difference
here is the explicit consideration of the truncation distance $D$.
Finally, it is worth noticing that all results derived below are exact for
the multi-type AST but an approximation for the original population
structure.

\section{Spreading dynamics} 
\label{sec:spreading}

To proceed further we should specify how the disease is transmitted from
an agent to its neighbors in the contact-graph. Let $r_{ab}$ be the
probability that an infected agent of type $a$ infects a susceptible
neighbor of type $b$. Within this work I assume that if $e_{ab}>0$ then
$r_{ab}>0$. Indeed, the absence of transmission between two types is taken 
into account by
the corresponding matrix element of $e$. Upon infection we also need to
specify when it takes place. Given a type $a$ agent (primary case)  and
one of its neighbors of type $b$ (secondary case), we define the
generation time $X_{ij}^{(a,b)}$ as the time elapse from the infection of
the primary case to the infection of the secondary case provided it
happens. I assume that the generation times are independent random
variables with the distribution function

\begin{equation}
G_{ab}(\tau) = {\rm Prob}\left( X_{ij}^{(a,b)}\leq\tau \right)\ ,
\label{Gij}
\end{equation}

\noindent parameterized by the type of the primary and secondary cases.

\begin{example}[SIR model]
\label{SIR}

In the SIR model agents can be in the three exclusive states susceptible,
infected and removed. A {\it susceptible} agent is one that have not
become infected but it is susceptible to acquire the infection. An {\it
infected} agent is one that have already acquired the disease and can
potentially transmit the disease. A {\it removed} agent is one that has
been previously infected but it is already excluded from the spreading
process. Within this work the removal of an agent takes into
account intervention strategies resulting in the isolation of infected
individuals from the disease transmission chain. The death or ``natural''
recovery of infected agents was already taken into account during the
definition of the contact-graph in subsection \ref{sec:cg}.

Consider an agent $i$ of type $a$ and one of its neighbors $j$ of type
$b$. Let $Y_{i,j}^{(a,b)}$ be the infection time of agent $j$ by $i$ in
the absence of agent's removal and let $G^{(a,b)}_{\rm I}(\tau) = {\rm
Prob}\left(Y_{i,j}^{(a,b)}\leq\tau \right)$ be its distribution function.  
Furthermore, let $Z_i^{(a)}$ be the removal time of agent $i$ in the
presence of agent's removal and let $G^{(a,b)}_{\rm R}(\tau) = {\rm
Prob}\left(Z_i^{(a)}\leq\tau \right)$ be its distribution function. The
probability that agent $j$ is infected by agent $i$ by time $t$ is given
by

\begin{equation}
b_{ab}(t) = \int_0^t dG_{\rm I}(\tau)\left[ 1- G_{\rm R}(\tau) \right]\ .
\label{bab}
\end{equation}

\noindent From this magnitude we obtain the probability that agent $j$
gets infected by agent $i$ no matter when

\begin{equation}
r_{ab} = \lim_{t\rightarrow\infty} b_{ab}(t)\ .
\label{rabSIR}
\end{equation}

\noindent and the distribution of generation times

\begin{equation}
G_{ab}(\tau) = \frac{1}{r_{ab}} b_{ab}(\tau)\ .
\label{GabSIR}
\end{equation}

The SIR model could be further generalized taking immunization into
account. In this case non-infected agents are divided into susceptible and
immune. If $s_a$ is the probability that a type $a$ agent is immune then
the probability that agent $j$ is infected by agent $i$ by time $t$ reads

\begin{equation}
b_{ab}(t) = (1-s_b)
\int_0^t dG_{\rm I}(\tau)\left[ 1- G_{\rm R}(\tau) \right]\ .
\label{babI}
\end{equation}

\noindent Furthermore, the transmission probability $r_{ab}$ and the 
generation time distribution $G_{ab}(\tau)$ are obtained substituting this 
equation into (\ref{rabSIR}) and (\ref{GabSIR}), respectively.

These examples illustrates how to calculate the transmission probability
$r_{ab}$ and the generation time distribution $G_{ab}(\tau)$ from the
standards models characterizing the spreading of infectious diseases. More
important, by encapsulating the model details into $r_{ab}$ and
$G_{ab}(\tau)$ we can obtain general results that are independent of these
details. Later on, we can analyze the particularities of each model.

\end{example}

\subsection{Multi-type age-dependent AST}

At this point the local spreading dynamics has been completely 
specified and we can super-impose it on the multi-type AST.

\begin{definition}{Multi-type age-dependent AST}

The multi-type age-dependent AST is composed of two elements, a multi-type
AST \ref{AST} and a local spreading dynamics defined by $\{r_{ab},\
G_{ab}(\tau)\}$.  The global dynamics is then specified by the following
rules

\begin{enumerate}

\item The process starts with an infected agent of type
$a\in\{1,\ldots,M\}$ while all other agents are susceptible.

\item An infected agent of type $a$ infects each of its neighbors of type
$b$ with probability $r_{ab}$ and generation time distribution
$G_{ab}(\tau)$.

\end{enumerate}

\label{adAST}
\end{definition}

\noindent The age-dependent AST is a generalization of the Bellman-Harris
\cite{harris02} and Crum-Mode-Jagers \cite{jagers75,mode00} multi-type
age-dependent branching processes. The key new element is the truncation
at a maximum generation, allowing us to consider the small-world property
of real networks. In spite of the similarities the mathematical framework I
implement deviates substantially from these previous approaches. Indeed, 
I exploit this truncation making a backward iteration from the final 
generation $D$ to the index case.

\section{Iterative approach}
\label{sec:iterative}

Consider a branch of the AST rooted on a type $a$ agent, at generation
$d$, that was infected at time zero. Let $P^{(d,a,b)}_n(t)$ be the
probability distribution to find $n$ infected type $b$ agents at time $t$
on that branch. In particular $P^{(0,a,b)}_n(t)$ is the probability
distribution of the total number of infected type $b$ agents at time $t$
on the whole AST, given the index case was of type $a$. Based on the tree
structure we can develop an iterative approach to compute
$P_n^{(d,a,b)}(t)$ recursively.

\begin{lemma}

Consider a type $a$ infected agent at generation $d$ of the multi-type
age-dependent AST. This agent has degree $k$ with probability $p^{(a)}_k$
for $d=0$ and excess degree $k-1$ with probability $q^{(a)}_k$ for
$0<d<D$. Let us index by $\alpha$ its neighbors on the next generation
$d+1$, where $\alpha\in\{1,\ldots,k\}$ for $d=0$,
$\alpha\in\{1,\ldots,k-1\}$ for $0<d<D$, and $\alpha\in\{\emptyset\}$ for
$d=D$. Then

\begin{eqnarray}
\displaystyle
P^{(0,a,b)}_n(t) & = & p^{(a)}_0 \left[ \delta_{ab}\delta_{n1}+
(1-\delta_{ab})\delta_{n0} \right]
\nonumber\\
& + & \sum_{k=1}^\infty p^{(a)}_k
\sum_{n_1=0}^\infty\dots\sum_{n_k=0}^\infty
\delta_{\sum_{\alpha=1}^k n_\alpha + \delta_{ab}, n}
\prod_{\alpha=1}^k \sum_{c=1}^M e_{ac}
\nonumber\\
& \times &
\left[ r_{ac}\int_0^t dG_{ac}(\tau) 
P^{(1,c,b)}_{n_\alpha}(t-\tau)
+ \delta_{n_\alpha,0}[ 1- r_{ac}G_{ac}(t)] \right]
\label{PN0}
\end{eqnarray}

\begin{eqnarray}
\displaystyle
P^{(d,a,b)}_n(t) & = & q^{(a)}_1 \left[ \delta_{ab}\delta_{n1}+
(1-\delta_{ab})\delta_{n0} \right]
\nonumber\\
& + & \sum_{k=2}^\infty q^{(a)}_k
\sum_{n_1=0}^\infty\dots\sum_{n_{k-1}=0}^\infty
\delta_{\sum_{\alpha=1}^{k-1}n_\alpha + \delta_{ab}, n}
\prod_{\alpha=1}^{k-1} \sum_{c=1}^M e_{ac}
\nonumber\\
& \times &
\left[ r_{ac}\int_0^t dG_{ac}(\tau) 
P^{(d+1,c,b)}_{n_\alpha}(t-\tau)
+ \delta_{n_\alpha,0}[ 1- r_{ac}G_{ac}(t)]\right]
\label{PNd}
\end{eqnarray}

\begin{equation}
P^{(D,a,b)}_n(t) = \delta_{ab}\delta_{n1}+
(1-\delta_{ab})\delta_{n0}\ .
\label{PND}
\end{equation}

\end{lemma}

\begin{proof}

Let $n$ be the number of infected type $b$ agents on a branch rooted at
type $a$ agent, and let $n_\alpha$ be the infected type $b$ agents on the
branches rooted at each of its neighbors $\alpha$.  Then

\begin{equation}
n = \delta_{ab} + \sum_\alpha n_\alpha\ ,
\label{NiNj}
\end{equation}

\noindent where $\delta_{ab}$ takes into account if the root agent is or 
it
is not of type $b$. The probability distribution of $n$ is given by the
sum of all the possible combinations of the random variables $n_{\alpha}$
that satisfy (\ref{NiNj}). Now, the root agent and its neighbors lie on a
tree and therefore $n_\alpha$ are independent random variables.
Furthermore, all agents at generation $d+1$ has the same statistical
properties, i.e. $n_\alpha$ are identically distributed random variables.
Therefore, the probability of each combination is decomposed into the
product of the probability distribution of the number of infected agents 
of
type $b$ on the sub-branches rooted at each neighbor. Thus, taking into
account that each neighbors is of type $c$ with probability
$e_{ac}$ we obtain

\begin{eqnarray}
P^{(0,a,b)}_n(t) & = & p^{(a)}_0 \left[ \delta_{ab}\delta_{n1}+
(1-\delta_{ab})\delta_{n0} \right]
\nonumber\\
& + & \sum_{k=1}^\infty p^{(a)}_k
\sum_{n_1=0}^\infty\dots\sum_{n_k=0}^\infty
\delta_{\sum_{\alpha=1}^kn_\alpha + \delta_{ab}, n}
\prod_{\alpha=1}^k
\sum_{c=1}^M e_{ac} Q_{n_\alpha}^{(d+1,a,c,b)}(t)\ ,
\label{PN0Q}
\end{eqnarray}

\begin{eqnarray}
P^{(d,a,b)}_n(t) & = & q^{(a)}_1 \left[ \delta_{ab}\delta_{n1}+
(1-\delta_{ab})\delta_{n0} \right]
\nonumber\\
& + & \sum_{k=2}^\infty q^{(a)}_k
\sum_{n_1=0}^\infty\dots\sum_{n_{k-1}=0}^\infty
\delta_{\sum_{\alpha=1}^{k-1} n_\alpha + \delta_{ab}, n}
\prod_{\alpha=1}^{k-1}
\sum_{c=1}^M e_{ac} Q_{n_\alpha}^{(d+1,a,c,b)}(t)\ ,
\label{PNdQ}
\end{eqnarray}

\noindent where $Q_{n_\alpha}^{(d+1,a,c,b)}(t)$ is the probability
distribution of $n_\alpha$ which we proceed to calculate. 

Let us focus on one neighbor and let us assume that it is of type $c$.
With probability $1-r_{ac}$ this agent is not infected at any time and
with probability $r_{ac}[1-G_{ac}(t)]$ it is not yet infected at time $t$
given it will be infected at some later time, resulting in

\begin{equation}
Q_0^{(d+1,a,c,b)}(t) = 1 -r_{ac}G_{ac}\ .
\label{Q0}
\end{equation}

\noindent On the other hand, with probability $r_{ac}$ the neighbor
is infected at some time $\tau$, with distribution function
$G_{ac}(\tau)$, and the spreading dynamics continue to subsequent
generations. Once the neighbor is infected the number of infected agents
of type $b$ on that sub-branch is a random variable with probability
distribution $P_n^{(d+1,c,b)}(t-\tau)$. Therefore, for $n>0$

\begin{equation}
Q_n^{(d+1,a,c,b)}(t) = r_{ac}
\int_0^tdG_{ac}(\tau)
P_n^{(d+1,c,b)}(t-\tau)\ .
\label{QN}
\end{equation}

\noindent Finally, substituting (\ref{Q0})  and (\ref{QN}) into
(\ref{PN0Q}) and (\ref{PNdQ}) we obtain equations (\ref{PN0}) and
(\ref{PNd}). The demonstration of (\ref{PND}) is straightforward. For
$d=D$ the process stops and therefore there is only one infected agent,
the root itself, which is or it is not of type $b$, resulting in
(\ref{PND}).

\end{proof}

Associated with the probability distribution $P^{(d,a,b)}_n(t)$
we introduce that generating function

\begin{equation}
F^{(d,a,b)}(x,t) = \sum_{n=0}^\infty P_n^{(d,a,b)}(t) x^n\ .
\label{Fdij}
\end{equation}

\noindent Using the recursive relations for the probability distribution
(\ref{PN0})-(\ref{PND}) we obtain the following recursive relations for
the generating function

\begin{equation}
F^{(0,a,b)}(x,t) = x^{\delta_{ab}} U_a \left( 
\sum_{c=1}^M e_{ac}
\left[ 1 - r_{ac} G_{ac}(t) +
r_{ac} \int_0^t dG_{ac}(\tau)
F^{(1,c,b)}(x,t-\tau) \right] \right)
\label{F0}
\end{equation}

\begin{equation}
F^{(d,a,b)}(x,t) = x^{\delta_{ab}} V_a \left( 
\sum_{c=1}^M e_{ac}
\left[ 1 - r_{ac} G_{ac}(t) +
r_{ac} \int_0^t dG_{ac}(\tau)
F^{(d,c,b)}(x,t-\tau) \right] \right)
\label{Fd}
\end{equation}

\begin{equation}
F^{(D,a,b)}(x,t) = x^{\delta_{ab}}\ .
\label{FD}
\end{equation}

\noindent These recursive equations are going to be useful in the
following calculations.

\section{Expected behavior}
\label{sec:average}

Given a infected agent of type $a$ the expected number of secondary 
infections of type $b$ it generates is given by

\begin{equation}
R_{ab} = \langle k\rangle_a e_{ab} r_{ab}
\label{Aij}
\end{equation}

\noindent if it is the index case and by

\begin{equation}
\tilde{R}_{ab} = \langle k\rangle_a^{\rm (excess)} e_{ab} r_{ab}
\label{Bij}
\end{equation}

\noindent otherwise. The matrices $R$ and $\tilde{R}$ are extensions of
the basic reproductive number to the multi-type case. In the following it
becomes clear that $\tilde{R}$ is more relevant and therefore I refer to
it as the reproductive number matrix.

\begin{lemma}

Consider an ensemble of multi-type age-dependent AST \ref{adAST} with
index case of type $a$. Let $N_{ab}(t)$ be the mean total number of
infected type $b$ agents at time $t$ and let $I_{ab}(t)dt$ be the mean
number of type $b$ agents that are infected between time $t$ and $t+dt$.
Then

\begin{equation}
N_{ab}(t) = \sum_{d=1}^D  \left(H\star 
J^{\star(d-1)}\right)_{ab}(t)\ ,
\label{NSijt}
\end{equation}

\begin{equation}
I_{ab}(t) = \sum_{d=1}^D  \frac{d}{dt}\left(H\star 
J^{\star(d-1)}\right)_{ab}(t)\ ,
\label{Iijt}
\end{equation}

\noindent where

\begin{equation}
H_{ab}(t) = R_{ab}G_{ab}(t)\ ,
\label{Hij}
\end{equation}

\begin{equation}
J_{ab}(t) = \tilde{R}_{ab}G_{ab}(t)\ ,
\label{Jij}
\end{equation}

\noindent and the multiplication symbolized by $\star$ involves a matrix
multiplication and a convolution in time. For instance,

\begin{equation}
\left(H\star J\right)_{ab}(t) = \sum_{c=1}^M \int_0^t d\tau
H_{ac}(\tau)J_{cb}(t-\tau)\ .
\label{ex1}
\end{equation} 

\begin{equation}
\left(J^{\star2}\right)_{ab}(t) = \left(J\star J\right)_{ab}(t)\ .
\label{ex2}
\end{equation} 

\label{ntgeneral}
\end{lemma}

\begin{proof}

\noindent Let

\begin{equation}
N^{(d,a,b)}(t) = \frac{\partial F^{(d,a,b)}(1,t)}{\partial x}
\label{Mdt}
\end{equation}

\noindent be the mean number of infected type $b$ agents on the branch
rooted at a type $a$ agent at generation $d$. In particular,
$N_{ab}(t)=N^{(0,a,b)}(t)$. Making use of the recursive relations
(\ref{F0})-(\ref{FD}) we obtain

\begin{equation}
N^{(0,a,b)}(t) = \delta_{ab} + \dot{U}_a(1) \sum_{c=1}^M
r_{ac}\int_0^tdG_{ac}(\tau)N^{(1,c,b)}(t-\tau)
\label{M0}
\end{equation}

\begin{equation}
N^{(d,a,b)}(t) = \delta_{ab} + \dot{V}_a(1) \sum_{c=1}^M
r_{ac}\int_0^tdG_{ac}(\tau)N^{(d+1,c,b)}(t-\tau)
\label{Md}
\end{equation}

\begin{equation}
N^{(D,a,b)}(t) = \delta_{ab}\ .
\label{MD}
\end{equation}

\noindent Iterating these recursive relations from $d=D$ to $d=0$ we
obtain (\ref{NSijt}). Then differentiating with respect to time we finally
obtain (\ref{Iijt}). In this step we also make use of the relation between
$\dot{U}(1)$ and $\dot{V}(1)$ and the average degrees
(\ref{Umeank})-(\ref{VmeanK}).

\end{proof}

\noindent This Lemma provides explicit equations for the expected
progression of an epidemic outbreak. In some particular cases these
equations may be further expressed in terms of elementary functions
allowing an straightforward interpretation. More generally these equations
can be evaluated numerically in cases where further reduction is not
possible. In addition, Theorem \ref{ntgeneral} is a starting point for
calculations addressing some limiting cases, which is the subject of the
following subsections.

\subsection{Final outbreak size}
\label{sec:finalsize}

The final outbreak size is obtained taking the limit $t\rightarrow\infty$
in (\ref{NSijt}), resulting in

\begin{equation}
N_{ab}(\infty) = \sum_{d=1}^D\left(R\tilde{R}^{d-1}\right)_{ab}\ .
\label{Noo}
\end{equation}

\noindent When $\tilde{R}$ can be diagonalized we can write
$\tilde{R}=PDP^{-1}$, where $P$ is the matrix composed of the eigenvectors
of $\tilde{R}$, $D$ is the diagonal matrix constructed from the
corresponding eigenvalues ($\rho_a$, $a=1,\ldots,M$) and $P^{-1}$ is the
inverse of $P$. Thus (\ref{Noo}) is reduced to

\begin{equation}
N_{ab}(\infty) = \left(RP\tilde{N}P^{-1}\right)_{ab}\ ,
\label{NooD}
\end{equation}

\noindent where $\tilde{N}$ is a diagonal matrix with diagonal entries

\begin{equation}
\tilde{N}_{aa} = \left\{
\begin{array}{ll}
\displaystyle
\frac{\rho_a^D-1}{\rho_a-1}\ , & \mbox{for}\ \rho_a\neq1\\
D\ , & \mbox{for}\ \rho_a=1
\end{array}
\right.
\label{tNaa}
\end{equation}

\noindent The following two Theorems show that the only thing we need to
estimate the order of magnitude of the expected outbreak size is the
largest eigenvalue of the reproductive number matrix $\tilde{R}$.

\begin{theorem}[Complete type-network]
\label{fsfm}

Consider a complete type-network and let $\rho$ be the largest
eigenvalue of $\tilde{R}$ (\ref{Bij}). Then

\begin{equation}
N_{ab}(\infty)= u_{ab} \frac{\rho^D-1}{\rho-1}\ , 
\label{Nijfm}
\end{equation}

\noindent where $u_{ab}$ is indenpendent of $D$.

\end{theorem}

\begin{proof}

The mixing matrix of a complete type-network is positive defined and,
therefore, $R$ (\ref{Aij}) and $\tilde{R}$ (\ref{Bij}) are positive
defined as well. From the Perron-Frobenius Theorem \cite{godsil01} it
follows that the largest eigenvalue of $\tilde{R}$ is simple and all the
entries of its corresponding left eigenvector $\vec{v}$ are different from
zero and have the same sign. In particular we choose all the components of
$\vec{v}$ to be positive such that

\begin{equation}
(R\tilde{R}^{d-1})_{ab} = \sum_{c=1}^M R_{ac} (\tilde{R}^{d-1})_{cb}
= \sum_{c=1}^M \frac{R_{ac}}{v_c} v_c(\tilde{R}^{d-1})_{cb}\ .
\label{Cij}
\end{equation}

\noindent Taking into account that $\sum_c v_c \tilde{R}_{cb} = \rho v_b$
we obtain the inequalities

\begin{equation}
u_{ab}^{\rm (min)}\rho^{d-1} \leq (R\tilde{R}^{d-1})_{ab} \leq
u_{ab}^{\rm (max)}\rho^{d-1}
\label{ineq1}
\end{equation}

\noindent where 

\begin{equation}
u_{ab}^{\rm (min)} = \min_{c}R_{ac}\frac{v_b}{v_c}\ ,
\label{aijmin}
\end{equation}

\begin{equation}
u_{ab}^{\rm (max)} = \max_{c}R_{ac}\frac{v_b}{v_c}\ ,
\label{aijmax}
\end{equation}

\noindent From (\ref{Nijfm}) and (\ref{Noo}) we obtain

\begin{equation}
1+u_{ab}^{\rm (min)}\frac{\rho^D-1}{\rho-1}
\leq N_{ab}(\infty) \leq
1+u_{ab}^{\rm (max)}\frac{\rho^D-1}{\rho-1}
\label{ineq2}
\end{equation}

\noindent Finally, from this equation we obtain (\ref{Nijfm}) 
with

\begin{equation}
0<u_{ab}^{\rm (min)} \leq u_{ab} \leq u_{ab}^{\rm (max)}<\infty\ ,
\label{AaA}
\end{equation}

\noindent where the inequality $u_{ab}>0$ follows from (\ref{aijmin}).

\end{proof}

\begin{figure}

\centerline{\includegraphics[height=3in]{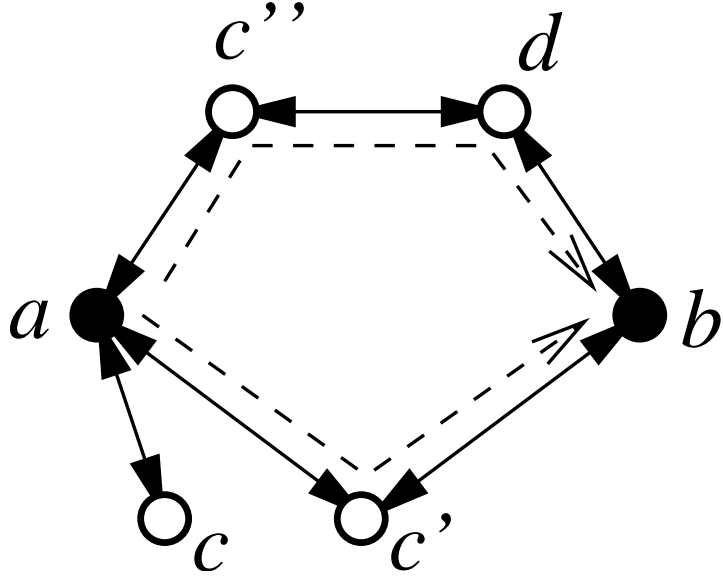}}

\caption{Strongly connected type-network with six types. The dashed lines
indicate the possible paths from type $a$ to $b$. Note that only the types
$c$, $c^\prime$ and $c^{\prime\prime}$ are neighbors of type $a$ and type
$b$ can be only reached from the last two.}

\label{fig:path}

\end{figure}

This result can be generalized to type-networks that may not be complete
but are still strongly connected, i.e. there is a path from every type
$a$ to every type $b$. In this case some entries of $R_{ac}$ and
$(\tilde{R}^{d-1})_{cb}$ in (\ref{Cij}) may be zero. Intuitively this
means that some types $c$ may not be a neighbor of $a$ and, if they are,
there may not be a path from $c$ to $b$ (See Fig. \ref{fig:path}). More
precisely, given a type $a$ let $Out(a)$ be its set of out-neighbors,
i.e. $Out(a)=\{c|e_{ac}>0\}$, and given a type $b$ let $In_d(b)$ be the
set of types from where $b$ is reached after $d$ hops on the type-network,
i.e.  $In_d(b)=\{c|(e^d)_{cb}>0\}$. Furthermore, let

\begin{equation}
S_d^{(a,b)}=Out(a)\cap In_{d-1}(b)
\label{S}
\end{equation}

\noindent denote the set of types that are out-neighbors of the index
case type $a$ and belong to at least one path of length $d$ from $a$ to
$b$ on the type-network. For instance, in the example in Fig.  
\ref{fig:path}, $S_1^{(a,b)}=\emptyset$, $S_2^{(a,b)}=\{c^\prime\}$,
$S_3^{(a,b)}=\{c^{\prime\prime}\}$, and $S_d^{(a,b)}\neq\emptyset$ for all 
$d>3$.

\begin{theorem}[Strongly connected type-network]
\label{fsc}

Consider a strongly connected type-network. Let $\rho$ be the largest
eigenvalue of $\tilde{R}$ (\ref{Bij}), $d_{ab}$ the distance on the
type-network from type $a$ to $b$, $n=[D/d_{ab}]$ and $D_{ab}=n d_{ab}$. Then

\begin{equation}
N_{ab}(\infty) = 
u_{ab} 
\sum_{1\leq d\leq D|S_d^{(a,b)}\neq\emptyset} \rho^{d-1}\ ,
\label{NijSC}
\end{equation}

\noindent where $u_{ab}$ is independent of $D$.

\end{theorem}

\begin{proof}

The conditions of the Perron-Frobenius theorem \cite{godsil01} are valid
beyond positive defined matrices and holds for the mixing matrix
representing a strongly connected network. Thus, the largest eigenvalue of
$\tilde{R}$ is simple and all the entries of its corresponding eigenvector
$\vec{v}$ are different from zero and have the same sign. In particular we
choose all the components of $\vec{v}$ to be positive. Based on this fact
we can write (\ref{Cij}). There may be, however, some entries of $e$ and
thus of $R$ (\ref{Aij}) and $\tilde{R}^{d-1}$ that are zero.  Indeed we
can rewrite (\ref{Cij}) as

\begin{equation}
(R\tilde{R}^{d-1})_{ab} = 
\sum_{1\leq c\leq M|c\in S_d^{(a,b)}}
\frac{R_{ac}}{v_c} v_c(\tilde{R}^{d-1})_{cb}\ .
\label{CijSC}
\end{equation}

\noindent Thus $(R\tilde{R}^{d-1})_{ab}=0$ whenever
$S_d^{(a,b)}=\emptyset$. Otherwise, we obtain the inequalities

\begin{equation}
u_{ab}^{\rm (min)}\rho^{d-1} \leq (R\tilde{R}^{d-1})_{ab} \leq
u_{ab}^{\rm (max)}\rho^{d-1}
\label{ineq1SC}
\end{equation}

\noindent where 

\begin{equation}
u_{ab}^{\rm (min)} = \min_{c\in Out(a)}\frac{v_b}{v_c}R_{ac}\ ,
\label{aijminSC}
\end{equation}

\begin{equation}
u_{ab}^{\rm (max)} = \max_{c\in Out(a)}\frac{v_b}{v_c}R_{ac}\ ,
\label{aijmaxSC}
\end{equation}

\noindent From (\ref{Nijfm}) and (\ref{Noo}) we obtain

\begin{equation}
1+u_{ab}^{\rm (min)} 
\sum_{1\leq d\leq D|S_d^{(a,b)}\neq\emptyset} \rho^{d-1}
\leq N_{ab}(\infty) \leq
1+u_{ab}^{\rm (max)}
\sum_{1\leq d\leq D|S_d^{(a,b)}\neq\emptyset} \rho^{d-1}
\label{ineq2SC}
\end{equation}

\noindent From this equation we obtain (\ref{NijSC})  with

\begin{equation}
0<u_{ab}^{\rm (min)} \leq u_{ab} \leq u_{ab}^{\rm (max)}<\infty\ ,
\label{AaASC}
\end{equation}

\noindent where the inequality $u_{ab}^{\rm (min)}>0$ follows from
(\ref{aijminSC}).

\end{proof}

\begin{figure}
\centerline{\includegraphics[height=3in]{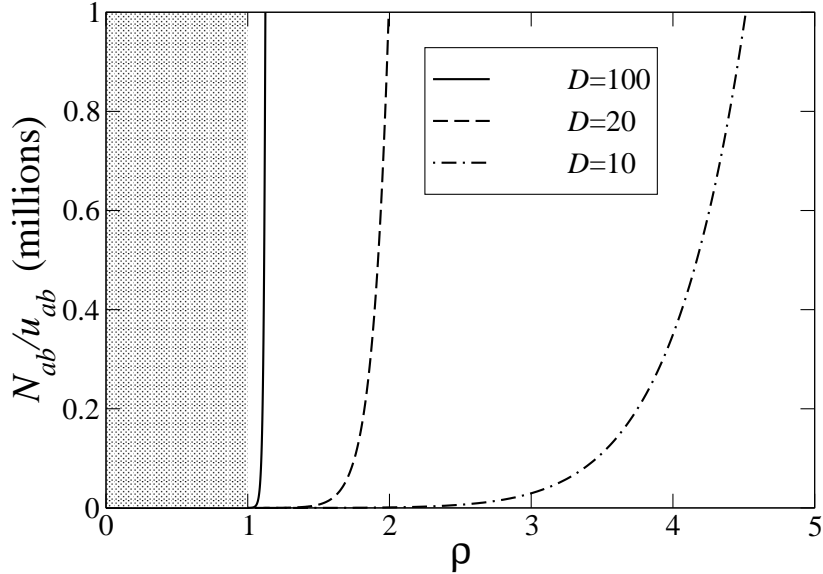}}

\caption{Expexted outbreak size as a function of the largest eigenvalue 
of the reproductive number matrix for different values of $D$. The 
region $\rho<1$ is indicated by the dotted pattern.}

\label{fig:Noo}
\end{figure}

Figure \ref{fig:Noo} illustrates the predictions of Theorem \ref{Nijfm}
for complete type-networks.  When $\rho<1$ the expected outbreak size is
of the order of the prefactor $u_{ab}$ which is not expected to be large.
Different behaviors are observed, however, for $\rho>1$ depending on $D$.
For $D\gg1$ there is a dramatic increase in the expected outbreak size. As
soon as $\rho>1$ a significant fraction of the agent population becomes
affected. In contrast, when $D$ is not so large it becomes clear that the
expected outbreak size changes smoothly with increasing $\rho$, including
the region around $\rho=1$. This fact becomes relevant when analyzing the
impact of intervention strategies (see section \ref{sec:intervention}).  
Finally, it is worth mentioning that a similar picture is obtained for the
more general case of strongly-connected type-networks, albeit some
corrections given by the missing terms the sum in (\ref{NijSC}).

\subsection{Spreading dynamics with constant transmission rate}

Now let us consider the particular case where the spreading dynamics is
homogeneous, i.e. $G_{ab}(\tau)=G(\tau)$. In this case, from (\ref{Iijt})
we obtain the incidence

\begin{equation}
I_{ab}(t) = \sum_{d=1}^D \left(R\tilde{R}^{d-1}\right)_{ab}
G^{\star d}(t)\ .
\label{Iijtht}
\end{equation}

\noindent In particular when $\tilde{R}$ can be diagonalized we rewrite
(\ref{Iijtht}) as

\begin{equation}
I_{ab}(t) = \left(RP\tilde{I}(t)P^{-1}\right)_{ab}\ ,
\label{ItD}
\end{equation}

\noindent where $\tilde{I}(t)$ is a time dependent diagonal matrix with
diagonal entries

\begin{equation}
\tilde{I}_{aa}(t) = \sum_{d=1}^D G^{\star d}(t)\ .
\label{tItD}
\end{equation}

\begin{example}
\label{sec:twocities}

Consider the case $M=2$ with the reproductive number matrices

\begin{equation}
R = \left[
\begin{matrix}
k_1 & k_2\\
k_2 & k_1\\
\end{matrix}
\right]\ ,\ \ \ \ 
\tilde{R} = \left[
\begin{matrix}
K_1 & K_2\\
K_2 & K_1\\
\end{matrix}
\right]
\label{tRtc}
\end{equation}

\noindent Since $\tilde{R}$ is symmetric it can be diagonalized and
$P^{-1}=P^{\rm T}$, where $P^{\rm T}$ is the transpose of $P$. In this
case $\tilde{R}=PDP^{\rm T}$ with

\begin{equation}
D = \left[
\begin{matrix}
\rho_1 & 0\\
0 & \rho_2\\
\end{matrix}
\right]\ ,\ \ \ \ 
P = \frac{1}{\sqrt{2}}\left[
\begin{matrix}
1 & 1\\
1 & -1\\
\end{matrix}
\right]
\label{Ptc}
\end{equation}

\noindent where

\begin{equation}
\rho=\rho_1 = K_1 + K_2\ ,\ \ \ \ 
\rho_2 = K_1 - K_2
\label{L2}
\end{equation}

\noindent are the eigenvalues of $\tilde{R}$. Assuming an index case
is of type $a=1$ from (\ref{ItD}) we finally obtain

\begin{equation}
I_{11}(t) = \frac{k_1+k_2}{2} \tilde{I}_{11}(t) + 
\frac{k_1-k_2}{2} \tilde{I}_{22}(t)
\label{I11}
\end{equation}

\begin{equation}
I_{12}(t) = \frac{k_1+k_2}{2} \tilde{I}_{11}(t) - 
\frac{k_1-k_2}{2} \tilde{I}_{22}(t)\ ,
\label{I12}
\end{equation}

\end{example}

This example shows that in some cases we can exactly calculate the
expected progression of an epidemic outbreak. More generally we obtain the
following asymptotic behaviors.

\begin{theorem}
\label{ntsc}

Consider a strongly connected type network and a homogeneous and
exponential distribution of generation times $G_{ab}=1-e^{-\lambda\tau}$,
where $\lambda$ is the transmission rate. Let $\rho$ be the largest
eigenvalue of $\tilde{R}$ (\ref{Bij}) and let

\begin{equation}
\theta = \frac{D-1}{\rho}\ .
\label{theta}
\end{equation}

$\theta\gg1$: If $\rho>1$  and $1\ll\lambda t\ll\theta$ then

\begin{equation}
I_{ab}(t) \sim e^{(\rho-1)\lambda t} \ .
\label{Ntexp}
\end{equation}

$\theta\ll1$: If $\lambda t\gg \theta$ then

\begin{equation}
\frac{I_{ab}(t)}{N_{ab}(\infty)} =
\frac{\lambda(\lambda t)^{D_{ab}-1}e^{-\lambda t}}{(D_{ab}-1)!} \left[
1 + {\cal O}\left( \frac{\theta}{\lambda t} \right) \right]\ ,
\label{Iijthtfc}
\end{equation}

\noindent where $D_{ab}$ is the same as in Theorem \ref{fsc}.

\end{theorem}

\begin{proof}

$\theta\gg1$: Following the same guidelines of the Theorem \ref{fsc} proof 
we arrive to the inequality

\begin{equation}
u_{ab}^{\rm (min)} f_{ab}(t)
\leq I_{ab}(t) \leq
u_{ab}^{\rm (max)} f_{ab}(t)\ ,
\label{ineq3}
\end{equation}

\noindent where

\begin{equation}
f_{ab}(t)= \sum_{1\leq d\leq D|S_d^{(a,b)}\neq\emptyset} 
\frac{\lambda(\rho\lambda t)^{d-1}e^{-\lambda t}}{(d-1)!}
\label{ft}
\end{equation}

\noindent The Laplace transform of $f_{ab}(t)$ is given by

\begin{equation}
\hat{f}_{ab}(\omega)= \int_0^\infty dt f_{ab}(t)e^{-\omega t} =
\frac{a}{\rho} \sum_{1\leq d\leq D|S_d^{(a,b)}\neq\emptyset} 
\left( \frac{\rho \lambda}{\omega + \lambda} \right)^d\ .
\label{fomega}
\end{equation}

\noindent When $D\rightarrow\infty$ this series converges only for
$\omega>(\rho -1)\lambda$. Therefore, $f_{ab}(t)\sim
e^{(\rho-1)\lambda t}$ when $\lambda t\rightarrow\infty$.

$\theta\ll1$: The demonstration of this case is straightforward.  From
Theorem \ref{fsc} it follows that $(R\tilde{R}^{d-1})_{ab}$ is of order
$\rho^{d-1}$ for $S_d^{(a,b)}\neq\emptyset$. Therefore, for $\rho\gg
D$ the sum in (\ref{Iijtht}) is dominated by the $d=D_{ab}$ term.
Corrections are given by the ratio between the $d=D_{ab}$ and the
preceeding term satisfying $S_d^{(a,b)}\neq\emptyset$, which is at most
$d=D_{ab}-1$.

\end{proof}

\noindent The case $\theta\gg1$ provides the connection between this work
and multi-type age-dependent branching processes with an infinite number
of generations. Indeed, Mode have already demonstrated the exponential
growth regime for the case $D=\infty$ (see \cite{mode71}, Chapter 3).
Theorem \ref{ntsc} shows that on the other limit $\theta\ll1$ the
spreading dynamics is instead characterized by a gamma distribution, which
is also the case for the single-type
case\cite{vazquez06a,vazquez06b,vazquez06e}.

\begin{figure}

\centerline{\includegraphics[width=5in]{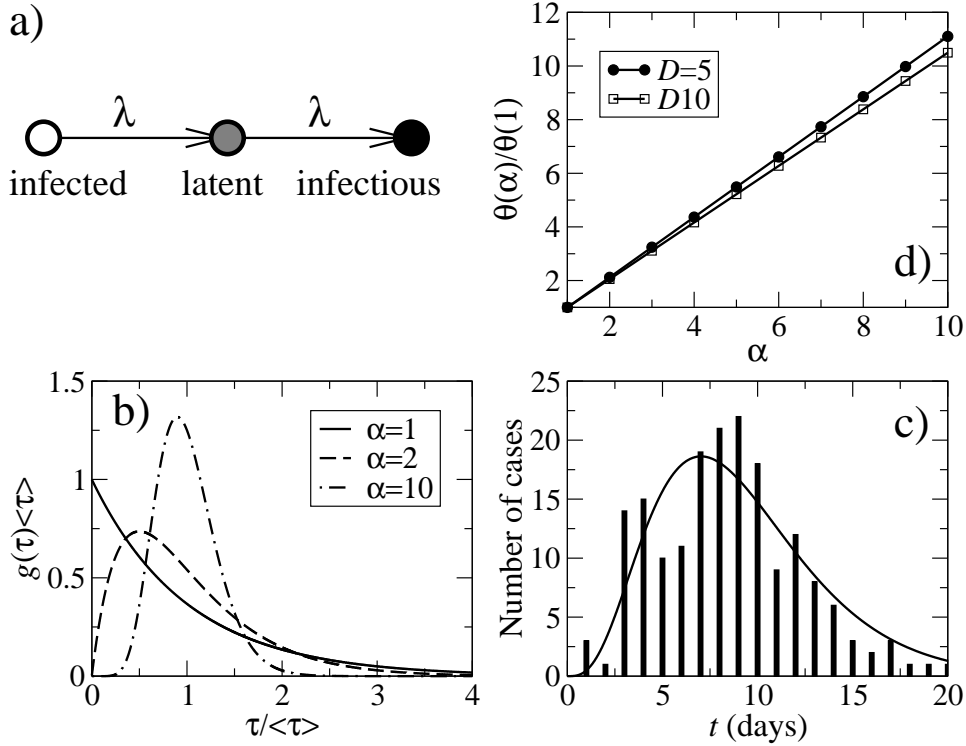}} 

\caption{(a) Schematic representation of the evolution of a disease within
an agent, starting from the moment the agent gets infected, passing
through a latent state where the agent is not infectious and finally
becoming infectious.  (b) Gamma probability density function
$g(\tau)=\dot{G}(\tau)$ for different values of $\alpha$. (c)  Number of
secondary cases generated by a primary case for a SARS outbreak in
Singapore, as reported in \cite{lipsitch03} (bars). The solid line is the
best fit to the gamma probability density function times a pre-factor,
resulting in $\alpha\approx4.3$. (d) Plot of the parameter
$\theta(\alpha)$ dividing the exponential and power law initial growth
regimes as a function of the number of intermediary steps.}

\label{fig:gtau}
\end{figure}

Theorem \ref{ntsc} can be extended to consider other generation time 
distributions, such as a gamma distribution

\begin{equation}
G(\tau) = \frac{1}{\Gamma(\alpha)}
\int_0^{\lambda t} dx x^{\alpha-1}e^{-x}\ ,
\label{gamma}
\end{equation}

\noindent where $\alpha\geq1$. The gamma distribution can be interpreted
as the existence of $\alpha-1$ intermediary steps before an agent becomes
infectious (see Fig.  \ref{fig:gtau}a,b). For $\alpha=1$ we recover the
exponential distribution which corresponds with the absence of
intermediary steps. The gamma distribution can be also obtained from the
fit to some empirical distribution of generation times (see Fig.  
\ref{fig:gtau}c).

In this case there are two important modifications to Theorem
\ref{ntsc}. First, the parameter $\theta$ is now given by

\begin{equation}
\theta(\alpha) = \frac{[(\alpha D-1)\cdots (\alpha D - \alpha)]
^{\frac{1}{\alpha}}}{\rho}\ ,
\label{thetaalpha}
\end{equation}

\noindent which increases approximately linearly with increasing $\alpha$
(see Fig. \ref{fig:gtau}d). Second, in the regime $\theta(\alpha)\gg1$
although the fraction of infected agents is still given by a gamma
distribution the exponent of the initial power law growth is given by
$\alpha D$, i.e.

\begin{equation}
\frac{I_{ab}(t)}{N_{ab}(\infty)} \approx
\frac{\lambda(\lambda t)^{\alpha D_{ab}-1}e^{-\lambda 
t}}{\Gamma(\alpha D_{ab}-1)}\ .
\label{Iijalpha}
\end{equation}

\noindent Therefore, the existence of intermediary steps reduces the the
small-world effect by a factor given by the number of intermediary steps
$\alpha$. For instance, by a factor of about four for SARS (Fig.
\ref{fig:gtau}b).

\subsection{Impact of intervention strategies}
\label{sec:intervention}

The expected outbreak size is a monotonic increasing function of $\rho$
(\ref{NijSC}), which plays the role of the basic reproductive number in
homogeneous populations \cite{anderson91,fraser04}. Therefore, the aim of
intervention strategies is to reduce the characteristic reproductive
number $\rho$. On the other hand, intervention strategies implies an
economical cost, including but not limited to the development of new
vaccines and their deployment through vaccination campaigns. Our task is
to design optimal intervention strategies that minimize the expected
outbreak size with a feasible economical cost.

To be more precise let us consider a scenario where the disease is
transmitted at constant rate $\lambda$ from infected to susceptible
agents, infected agents are isolated at a rate $\mu$ and a fraction $s$ of
the population is immune to the disease. In this case the infection and
removal times follows the exponential distribution functions $G_{\rm
I}(\tau)=1-e^{-\lambda\tau}$ and $G_{\rm R}(\tau)=1-e^{-\mu\tau}$,
respectively. Thus, from (\ref{rabSIR}) we obtain $r_{ab}=1-\beta$ where

\begin{equation}
\beta = 1- \frac{\lambda}{\lambda+\mu}(1-s)\ ,
\label{thetas}
\end{equation}

\noindent is the blocking fraction, i.e. the fraction of potential disease
transmissions that are blocked either because of immunization or patient
isolation. Since $r_{ab}=1-\beta$ is independent of the
primary and secondary case types we can write the reproductive number
matrices (\ref{Aij}) and (\ref{Bij}) as $R_{ab}=(1-\beta)K$ and
$\tilde{R}=(1-\beta)\tilde{K}$, respectively, where

\begin{equation}
K_{ab}=\langle k\rangle_a e_{ab}\ ,\ \ \ \
\tilde{K}_{ab}=\langle k\rangle_a^{(\rm excess)} e_{ab}\ . 
\label{K}
\end{equation}

\noindent In turn, the largest eigenvalue of $\tilde{R}$ is given by

\begin{equation}
\rho = (1-\beta) \kappa\ ,
\label{lk}
\end{equation}

\noindent where $\kappa$ is the largest eigenvalue of $\tilde{K}$.

From the analysis made in section \ref{sec:finalsize} it follows that
there are two different scenarios depending $D_{ab}$. For simplicity let
us focus on the complete type-network case where $D_{ab}=D$. When $D\gg1$
the target of intervention strategies is $\rho=1$, which is the consensus
in the literature \cite{anderson91,fraser04}. The blocking fraction to
achieve this is obtained from (\ref{lk}), resulting in

\begin{equation}
\beta_c = 1-\frac{1}{\kappa}\ .
\label{thetac}
\end{equation} 

\noindent This result has been already reported, at least for the case of
two types \cite{anderson91}.  When $D$ is small, however, the expected
outbreak size is a smooth function of $\rho$ (see Fig. \ref{fig:Noo}).
Therefore, $\beta_c$ does not represent a threshold value in small-world
populations.

So far we have considered homogenous intervention strategies. Now let us
assume that the rate of patient isolation and the immunized fraction are
now different for each agent's type and given by $\mu_a$ and $s_a$,
respectively. In this case the blocking fraction is given by

\begin{equation}
\beta_{ab} = 1- \frac{\lambda}{\lambda+\mu_a}(1-s_b)\ ,
\label{thetaab}
\end{equation}

\noindent and $r_{ab}=1-\beta_{ab}$, which depends on the type of both the
primary and secondary case. From the Perron-Frobenius Theorem it follows
that $\rho$ is a continuous increasing function of all the entries of the
corresponding matrix $\tilde{R}$ \cite{gantmacher90}. Since
$\tilde{R}_{ab}=(1-\beta_{ab})\tilde{K}_{ab}$ then $\rho$ is a continuous
decreasing function of $\theta_{ab}$ for all $(a,b)$. The goal is to
determine which strategy leads to the largest reduction of $\rho$.

\begin{example}

Consider the spread of HIV on an heterosexual population with no further
distinction beyond gender. In this case the type-network is bipartite (see
Fig. \ref{typegraph}b). Let $k_1$ and $k_2$ be the average excess degree
for the connections from women to men and biceversa. Let also assume that
the rate of patient isolation is zero and that we could immunize a
fraction $s$ of the overall population, distributed between a fraction
$xs$ and $(1-x)s$ of immunized women and men, respectively. The question
is to determine the value of $x$ representing the best intervention
strategy. In this case the reproductive number matrix is given by

\begin{equation}
\tilde{R} = \left[
\begin{matrix}
0 & [1-(1-x)s]k_2\\
(1-xs)k_1 & 0
\end{matrix}
\right]
\label{tRb}
\end{equation}

\noindent and it has the largest eigenvalue

\begin{equation}
\rho = \sqrt{[1-s+x(1-x)s^2] k_1 k_2}\ .
\label{lb}
\end{equation}

\noindent It results that $\rho$ is minimum for $x=0$ or $x=1$, i.e. the
best intervention strategy is to direct all the immunization resources to
only one of the sub-populations.

\end{example}

\section{Discussion}
\label{sec:discussion}

There is significant evidence that social networks are characterized by
({\it i}) wide connectivity fluctuations and ({\it ii}) the small-world
property \cite{ws98}. The variability in the number of contacts ({\it i})
has a direct impact on the reproductive number. This fact has been taken
into account since the seminal works of May and Anderson considering the
variability in the rate of sexual partner
acquisition\cite{may87,may88,anderson91}. More recently it has gained
attention for other infectious diseases as well, following the observation
of super-spreading events in the 2002-2003 SARS epidemics
\cite{anderson04,galvani05,lloyd05}. Yet, the small-world property ({\it
ii}) has been completely neglected.

From my studies of the single type case
\cite{vazquez06a,vazquez06b,vazquez06e} I have shown that intervention
strategies are modulated by the average distance $D$ between agents in the
corresponding contact-graph. In this work I have demonstrated that this
result is also valid for heterogeneous populations. In this last case the
characteristic reproductive number is given by the largest eigenvalue of
the reproductive number matrix.  The good news is that in spite of this
modulation by $D$ the target of intervention strategies is still the
characteristic reproductive number.  That is, the expected outbreak size
still decreases with decreasing the characteristic reproductive number.
The bad news is that to quantify the impact of the intervention strategies
we need to estimate $D$.

There are different paths to estimate $D$. First, we can use a direct
approach as the Milgram's experiments \cite{m67}. Second, we can
measure other network properties such as the degree distribution and then
try to estimate $D$ using network models
\cite{bollobas01,chung02,bollobas03,cohen03a,vazquez03a,leskovec06}.  
Finally, I have shown that the progression of the expected number of new
infections is modulated by $D$ (see
\cite{vazquez06a,vazquez06b,vazquez06e} and section \ref{sec:spreading}).  
More precissely, in small world populations the incidence is expected to
grow as a power law and we can estimate $D$ from the power law exponent.

Further work is required to test the validity of the coarse grained 
description of the type-network approach. This can be done by running 
agent based simulations where we can have a strict control of the 
different statistical properties characterizing the population structure. 
These statistical properties can be then plug in into the type network 
approach to obtain qualitative and quantitative predictions that can be 
compared with the simulations results. 

In conclusion, this work opens new avenues to future research on the
spreading of infectious diseases on heterogeneous populations. It allows
for a qualitative understanding through the analysis of the type-network
representation of the mixing matrix. More important, it leads to general
results that can be tackled case by case using exact or approximate
calculations and numerical computations.

This work was supported by NSF Grants No. ITR 0426737 and No. ACT/SGER 
0441089.

%\bibliography{network}

\begin{thebibliography}{48}
\expandafter\ifx\csname natexlab\endcsname\relax\def\natexlab#1{#1}\fi
\expandafter\ifx\csname bibnamefont\endcsname\relax
  \def\bibnamefont#1{#1}\fi
\expandafter\ifx\csname bibfnamefont\endcsname\relax
  \def\bibfnamefont#1{#1}\fi
\expandafter\ifx\csname citenamefont\endcsname\relax
  \def\citenamefont#1{#1}\fi
\expandafter\ifx\csname url\endcsname\relax
  \def\url#1{\texttt{#1}}\fi
\expandafter\ifx\csname urlprefix\endcsname\relax\def\urlprefix{URL }\fi
\providecommand{\bibinfo}[2]{#2}
\providecommand{\eprint}[2][]{\url{#2}}

\bibitem[{\citenamefont{Anderson and May}(1991)}]{anderson91}
\bibinfo{author}{\bibfnamefont{R.~M.} \bibnamefont{Anderson}} \bibnamefont{and}
  \bibinfo{author}{\bibfnamefont{R.~M.} \bibnamefont{May}},
  \emph{\bibinfo{title}{Infectious diseases of humans}}
  (\bibinfo{publisher}{Oxford Univ. Press, New York}, \bibinfo{year}{1991}).

\bibitem[{\citenamefont{Hufnagel et~al.}(2004)\citenamefont{Hufnagel,
  Brockmann, and T}}]{hufnagel04}
\bibinfo{author}{\bibfnamefont{L.}~\bibnamefont{Hufnagel}},
  \bibinfo{author}{\bibfnamefont{D.}~\bibnamefont{Brockmann}},
  \bibnamefont{and} \bibinfo{author}{\bibfnamefont{G.}~\bibnamefont{T}},
  \bibinfo{journal}{Proc. Natl. Acad. Sci. USA} \textbf{\bibinfo{volume}{101}},
  \bibinfo{pages}{15124} (\bibinfo{year}{2004}).

\bibitem[{\citenamefont{Koopman}(2004)}]{koopman04}
\bibinfo{author}{\bibfnamefont{J.}~\bibnamefont{Koopman}},
  \bibinfo{journal}{Annu. Rev. Public Health} \textbf{\bibinfo{volume}{25}},
  \bibinfo{pages}{303} (\bibinfo{year}{2004}).

\bibitem[{\citenamefont{Germann et~al.}(2006)\citenamefont{Germann, Kadau,
  Longini, and Macken}}]{germann06}
\bibinfo{author}{\bibfnamefont{T.~C.} \bibnamefont{Germann}},
  \bibinfo{author}{\bibfnamefont{K.}~\bibnamefont{Kadau}},
  \bibinfo{author}{\bibfnamefont{I.~M.} \bibnamefont{Longini}},
  \bibnamefont{and} \bibinfo{author}{\bibfnamefont{C.~A.}
  \bibnamefont{Macken}}, \bibinfo{journal}{Proc. Natl. Acad. Sci. USA}
  \textbf{\bibinfo{volume}{103}}, \bibinfo{pages}{5935} (\bibinfo{year}{2006}).

\bibitem[{\citenamefont{Colizza et~al.}(2006)\citenamefont{Colizza, Barrat,
  Barthelemy, and Vespignani}}]{colizza06}
\bibinfo{author}{\bibfnamefont{V.}~\bibnamefont{Colizza}},
  \bibinfo{author}{\bibfnamefont{A.}~\bibnamefont{Barrat}},
  \bibinfo{author}{\bibfnamefont{M.}~\bibnamefont{Barthelemy}},
  \bibnamefont{and}
  \bibinfo{author}{\bibfnamefont{A.}~\bibnamefont{Vespignani}},
  \bibinfo{journal}{Proc. Natl. Acad. Sci. USA} \textbf{\bibinfo{volume}{103}},
  \bibinfo{pages}{2015} (\bibinfo{year}{2006}).

\bibitem[{\citenamefont{May and Anderson}(1987)}]{may87}
\bibinfo{author}{\bibfnamefont{R.~M.} \bibnamefont{May}} \bibnamefont{and}
  \bibinfo{author}{\bibfnamefont{R.~M.} \bibnamefont{Anderson}},
  \bibinfo{journal}{Nature} \textbf{\bibinfo{volume}{326}},
  \bibinfo{pages}{137} (\bibinfo{year}{1987}).

\bibitem[{\citenamefont{Anderson and {\it et al}}(2004)}]{anderson04}
\bibinfo{author}{\bibfnamefont{R.~M.} \bibnamefont{Anderson}} \bibnamefont{and}
  \bibinfo{author}{\bibnamefont{{\it et al}}}, \bibinfo{journal}{Phil. Trans.
  R. Soc. Lond. B} \textbf{\bibinfo{volume}{359}}, \bibinfo{pages}{1091}
  (\bibinfo{year}{2004}).

\bibitem[{\citenamefont{Lloyd-Smith et~al.}(2005)\citenamefont{Lloyd-Smith,
  Schreiber, Kopp, and Getz}}]{lloyd05}
\bibinfo{author}{\bibfnamefont{J.~O.} \bibnamefont{Lloyd-Smith}},
  \bibinfo{author}{\bibfnamefont{S.~J.} \bibnamefont{Schreiber}},
  \bibinfo{author}{\bibfnamefont{P.~E.} \bibnamefont{Kopp}}, \bibnamefont{and}
  \bibinfo{author}{\bibfnamefont{W.~M.} \bibnamefont{Getz}},
  \bibinfo{journal}{Nature} \textbf{\bibinfo{volume}{438}},
  \bibinfo{pages}{355} (\bibinfo{year}{2005}).

\bibitem[{\citenamefont{May and Anderson}(1988)}]{may88}
\bibinfo{author}{\bibfnamefont{T.~M.} \bibnamefont{May}} \bibnamefont{and}
  \bibinfo{author}{\bibfnamefont{R.~M.} \bibnamefont{Anderson}},
  \bibinfo{journal}{Phil. Trans. R. Soc. Lond. B}
  \textbf{\bibinfo{volume}{321}}, \bibinfo{pages}{565} (\bibinfo{year}{1988}).

\bibitem[{\citenamefont{Liljeros et~al.}(2001)\citenamefont{Liljeros, Edling,
  Amaral, Stanley, and Berg}}]{liljeros01}
\bibinfo{author}{\bibfnamefont{F.}~\bibnamefont{Liljeros}},
  \bibinfo{author}{\bibfnamefont{C.~R.} \bibnamefont{Edling}},
  \bibinfo{author}{\bibfnamefont{L.~A.~N.} \bibnamefont{Amaral}},
  \bibinfo{author}{\bibfnamefont{H.~E.} \bibnamefont{Stanley}},
  \bibnamefont{and} \bibinfo{author}{\bibfnamefont{Y.}~\bibnamefont{Berg}},
  \bibinfo{journal}{Nature} \textbf{\bibinfo{volume}{411}},
  \bibinfo{pages}{907} (\bibinfo{year}{2001}).

\bibitem[{\citenamefont{Jones and Handcock}(2003)}]{jones03}
\bibinfo{author}{\bibfnamefont{J.~H.} \bibnamefont{Jones}} \bibnamefont{and}
  \bibinfo{author}{\bibfnamefont{M.~S.} \bibnamefont{Handcock}},
  \bibinfo{journal}{Proc. R. Soc. Lond. B Biol. Sci.}
  \textbf{\bibinfo{volume}{270}}, \bibinfo{pages}{1123} (\bibinfo{year}{2003}).

\bibitem[{\citenamefont{Schneeberger et~al.}(2004)\citenamefont{Schneeberger,
  Mercer, Gregson, Fergurson, Nyamukapa, Anderson, Johnson, and
  Garnett}}]{schneeberger04}
\bibinfo{author}{\bibfnamefont{A.}~\bibnamefont{Schneeberger}},
  \bibinfo{author}{\bibfnamefont{C.~H.} \bibnamefont{Mercer}},
  \bibinfo{author}{\bibfnamefont{S.~A.} \bibnamefont{Gregson}},
  \bibinfo{author}{\bibfnamefont{N.~M.} \bibnamefont{Fergurson}},
  \bibinfo{author}{\bibfnamefont{C.~A.} \bibnamefont{Nyamukapa}},
  \bibinfo{author}{\bibfnamefont{R.~M.} \bibnamefont{Anderson}},
  \bibinfo{author}{\bibfnamefont{A.~M.} \bibnamefont{Johnson}},
  \bibnamefont{and} \bibinfo{author}{\bibfnamefont{G.~P.}
  \bibnamefont{Garnett}}, \bibinfo{journal}{Sex. Transm. Dis.}
  \textbf{\bibinfo{volume}{31}}, \bibinfo{pages}{380} (\bibinfo{year}{2004}).

\bibitem[{\citenamefont{Pastor-Satorras and Vespignani}(2001)}]{pv00}
\bibinfo{author}{\bibfnamefont{R.}~\bibnamefont{Pastor-Satorras}}
  \bibnamefont{and}
  \bibinfo{author}{\bibfnamefont{A.}~\bibnamefont{Vespignani}},
  \bibinfo{journal}{Phys. Rev. Lett.} \textbf{\bibinfo{volume}{86}},
  \bibinfo{pages}{3200} (\bibinfo{year}{2001}).

\bibitem[{\citenamefont{Moore and Newman}(2000)}]{mn00}
\bibinfo{author}{\bibfnamefont{C.}~\bibnamefont{Moore}} \bibnamefont{and}
  \bibinfo{author}{\bibfnamefont{M.~E.~J.} \bibnamefont{Newman}},
  \bibinfo{journal}{Phys. Rev. E} \textbf{\bibinfo{volume}{61}},
  \bibinfo{pages}{5678} (\bibinfo{year}{2000}).

\bibitem[{\citenamefont{Albert and Barab\'asi}(2001)}]{ab01a}
\bibinfo{author}{\bibfnamefont{R.}~\bibnamefont{Albert}} \bibnamefont{and}
  \bibinfo{author}{\bibfnamefont{A.-L.} \bibnamefont{Barab\'asi}},
  \bibinfo{journal}{Rev. Mod. Phys.} \textbf{\bibinfo{volume}{74}},
  \bibinfo{pages}{47} (\bibinfo{year}{2001}).

\bibitem[{\citenamefont{Rvachev and Longini}(1985)}]{rvachev85}
\bibinfo{author}{\bibfnamefont{L.~A.} \bibnamefont{Rvachev}} \bibnamefont{and}
  \bibinfo{author}{\bibfnamefont{I.~M.} \bibnamefont{Longini}},
  \bibinfo{journal}{Math. Biosci.} \textbf{\bibinfo{volume}{75}},
  \bibinfo{pages}{3} (\bibinfo{year}{1985}).

\bibitem[{\citenamefont{Flahault et~al.}(1988)\citenamefont{Flahault, Letrait,
  Blin, Hazout, Menares, and Valleron}}]{flahault88}
\bibinfo{author}{\bibfnamefont{A.}~\bibnamefont{Flahault}},
  \bibinfo{author}{\bibfnamefont{S.}~\bibnamefont{Letrait}},
  \bibinfo{author}{\bibfnamefont{P.}~\bibnamefont{Blin}},
  \bibinfo{author}{\bibfnamefont{S.}~\bibnamefont{Hazout}},
  \bibinfo{author}{\bibfnamefont{J.}~\bibnamefont{Menares}}, \bibnamefont{and}
  \bibinfo{author}{\bibfnamefont{A.~J.} \bibnamefont{Valleron}},
  \bibinfo{journal}{Stat. Med.} \textbf{\bibinfo{volume}{7}},
  \bibinfo{pages}{1147} (\bibinfo{year}{1988}).

\bibitem[{\citenamefont{Eubank et~al.}(2004)\citenamefont{Eubank, Guclu, Kumar,
  Marathe, Srinivasan, Toroczcai, and Wang}}]{eubank04}
\bibinfo{author}{\bibfnamefont{S.}~\bibnamefont{Eubank}},
  \bibinfo{author}{\bibfnamefont{H.}~\bibnamefont{Guclu}},
  \bibinfo{author}{\bibfnamefont{V.~S.~A.} \bibnamefont{Kumar}},
  \bibinfo{author}{\bibfnamefont{M.}~\bibnamefont{Marathe}},
  \bibinfo{author}{\bibfnamefont{A.}~\bibnamefont{Srinivasan}},
  \bibinfo{author}{\bibfnamefont{Z.}~\bibnamefont{Toroczcai}},
  \bibnamefont{and} \bibinfo{author}{\bibfnamefont{N.}~\bibnamefont{Wang}},
  \bibinfo{journal}{Nature} \textbf{\bibinfo{volume}{429}},
  \bibinfo{pages}{180} (\bibinfo{year}{2004}).

\bibitem[{\citenamefont{Newman}(2003)}]{newman03}
\bibinfo{author}{\bibfnamefont{M.~E.~J.} \bibnamefont{Newman}},
  \bibinfo{journal}{Phys. Rev. E} \textbf{\bibinfo{volume}{67}},
  \bibinfo{pages}{026126} (\bibinfo{year}{2003}).

\bibitem[{\citenamefont{Mode}(1971)}]{mode71}
\bibinfo{author}{\bibfnamefont{C.~J.} \bibnamefont{Mode}},
  \emph{\bibinfo{title}{Multitype branching processes}}
  (\bibinfo{publisher}{Elsevier, New York}, \bibinfo{year}{1971}).

\bibitem[{\citenamefont{Vazquez}(2006{\natexlab{a}})}]{vazquez06a}
\bibinfo{author}{\bibfnamefont{A.}~\bibnamefont{Vazquez}}, in
  \emph{\bibinfo{booktitle}{AMS-DIMACS Volume on Discrete Methods in
  Epidemiology}} (\bibinfo{publisher}{AMS, in press},
  \bibinfo{year}{2006}{\natexlab{a}}).

\bibitem[{\citenamefont{Vazquez}(2006{\natexlab{b}})}]{vazquez06b}
\bibinfo{author}{\bibfnamefont{A.}~\bibnamefont{Vazquez}},
  \bibinfo{journal}{Phys. Rev. Lett.} \textbf{\bibinfo{volume}{96}},
  \bibinfo{pages}{038702} (\bibinfo{year}{2006}{\natexlab{b}}).

\bibitem[{\citenamefont{Vazquez}()}]{vazquez06e}
\bibinfo{author}{\bibfnamefont{A.}~\bibnamefont{Vazquez}},
  \bibinfo{note}{http://arxiv.org/q-bio.PE/0603010}.

\bibitem[{\citenamefont{Friedman and {\it et al}}(1997)}]{friedman97}
\bibinfo{author}{\bibfnamefont{S.~R.} \bibnamefont{Friedman}} \bibnamefont{and}
  \bibinfo{author}{\bibnamefont{{\it et al}}}, \bibinfo{journal}{Am. J. Pub.
  Health} \textbf{\bibinfo{volume}{87}}, \bibinfo{pages}{1289}
  (\bibinfo{year}{1997}).

\bibitem[{\citenamefont{Edmunds et~al.}(1997)\citenamefont{Edmunds,
  O'Callaghan, and Nokes}}]{edmunds97}
\bibinfo{author}{\bibfnamefont{W.~J.} \bibnamefont{Edmunds}},
  \bibinfo{author}{\bibfnamefont{C.~J.~O.} \bibnamefont{O'Callaghan}},
  \bibnamefont{and} \bibinfo{author}{\bibfnamefont{D.~J.} \bibnamefont{Nokes}},
  \bibinfo{journal}{Proc. R. Soc. Lond. B} \textbf{\bibinfo{volume}{264}},
  \bibinfo{pages}{949} (\bibinfo{year}{1997}).

\bibitem[{\citenamefont{Ghani and Garnett}(1998)}]{ghani98}
\bibinfo{author}{\bibfnamefont{A.~C.} \bibnamefont{Ghani}} \bibnamefont{and}
  \bibinfo{author}{\bibfnamefont{G.~P.} \bibnamefont{Garnett}},
  \bibinfo{journal}{J. R. Statist. Soc. A} \textbf{\bibinfo{volume}{161}},
  \bibinfo{pages}{227} (\bibinfo{year}{1998}).

\bibitem[{\citenamefont{Keeling and Eames}(2005)}]{keeling05}
\bibinfo{author}{\bibfnamefont{M.~J.} \bibnamefont{Keeling}} \bibnamefont{and}
  \bibinfo{author}{\bibfnamefont{K.~T.~D.} \bibnamefont{Eames}},
  \bibinfo{journal}{J. R. Soc. Interface} \textbf{\bibinfo{volume}{2}},
  \bibinfo{pages}{297} (\bibinfo{year}{2005}).

\bibitem[{\citenamefont{Watts and May}(1992)}]{watts92}
\bibinfo{author}{\bibfnamefont{C.~H.} \bibnamefont{Watts}} \bibnamefont{and}
  \bibinfo{author}{\bibfnamefont{R.~M.} \bibnamefont{May}},
  \bibinfo{journal}{Math. Biosci.} \textbf{\bibinfo{volume}{108}},
  \bibinfo{pages}{89} (\bibinfo{year}{1992}).

\bibitem[{\citenamefont{Kretzschmar and Morris}(1996)}]{kretzschmar96}
\bibinfo{author}{\bibfnamefont{M.}~\bibnamefont{Kretzschmar}} \bibnamefont{and}
  \bibinfo{author}{\bibfnamefont{M.}~\bibnamefont{Morris}},
  \bibinfo{journal}{Math. Biosci.} \textbf{\bibinfo{volume}{133}},
  \bibinfo{pages}{165} (\bibinfo{year}{1996}).

\bibitem[{\citenamefont{Garnett and A.~M}(1997)}]{garnett97}
\bibinfo{author}{\bibfnamefont{G.~P.} \bibnamefont{Garnett}} \bibnamefont{and}
  \bibinfo{author}{\bibfnamefont{J.}~\bibnamefont{A.~M}},
  \bibinfo{journal}{AIDS} \textbf{\bibinfo{volume}{11}}, \bibinfo{pages}{681}
  (\bibinfo{year}{1997}).

\bibitem[{\citenamefont{Watts}(1999)}]{w99}
\bibinfo{author}{\bibfnamefont{D.~J.} \bibnamefont{Watts}},
  \emph{\bibinfo{title}{Small Worlds: The Dynamics of Networks between Order
  and Randomness}} (\bibinfo{publisher}{Princeton University Press, Princeton,
  New Jersey}, \bibinfo{year}{1999}).

\bibitem[{\citenamefont{Milgram}(1967)}]{m67}
\bibinfo{author}{\bibfnamefont{S.}~\bibnamefont{Milgram}},
  \bibinfo{journal}{Psychology today} \textbf{\bibinfo{volume}{2}},
  \bibinfo{pages}{60} (\bibinfo{year}{1967}).

\bibitem[{\citenamefont{Watts and Strogatz}(1998)}]{ws98}
\bibinfo{author}{\bibfnamefont{D.~J.} \bibnamefont{Watts}} \bibnamefont{and}
  \bibinfo{author}{\bibfnamefont{S.~H.} \bibnamefont{Strogatz}},
  \bibinfo{journal}{Nature} \textbf{\bibinfo{volume}{393}},
  \bibinfo{pages}{440} (\bibinfo{year}{1998}).

\bibitem[{\citenamefont{Bollob\'{a}s}(2001)}]{bollobas01}
\bibinfo{author}{\bibfnamefont{B.}~\bibnamefont{Bollob\'{a}s}},
  \emph{\bibinfo{title}{Random Graphs}} (\bibinfo{publisher}{Cambridge:
  Cambridge University Press}, \bibinfo{year}{2001}).

\bibitem[{\citenamefont{Chung and Lu}(2002)}]{chung02}
\bibinfo{author}{\bibfnamefont{F.}~\bibnamefont{Chung}} \bibnamefont{and}
  \bibinfo{author}{\bibfnamefont{L.}~\bibnamefont{Lu}}, \bibinfo{journal}{Proc.
  Natl. Acad. Sci. USA} \textbf{\bibinfo{volume}{99}}, \bibinfo{pages}{15879}
  (\bibinfo{year}{2002}).

\bibitem[{\citenamefont{Bollob\'as and Riordan}(2003)}]{bollobas03}
\bibinfo{author}{\bibfnamefont{B.}~\bibnamefont{Bollob\'as}} \bibnamefont{and}
  \bibinfo{author}{\bibfnamefont{O.~M.} \bibnamefont{Riordan}}, in
  \emph{\bibinfo{booktitle}{Handbook of Graphs and Networks: From the Genome to
  the Internet}}, edited by
  \bibinfo{editor}{\bibfnamefont{S.}~\bibnamefont{Bornholdt}} \bibnamefont{and}
  \bibinfo{editor}{\bibfnamefont{H.~G.} \bibnamefont{Schuster}}
  (\bibinfo{publisher}{Wiley-VCH, Weinheim}, \bibinfo{year}{2003}), pp.
  \bibinfo{pages}{1--34}.

\bibitem[{\citenamefont{Cohen and Havlin}(2003)}]{cohen03a}
\bibinfo{author}{\bibfnamefont{R.}~\bibnamefont{Cohen}} \bibnamefont{and}
  \bibinfo{author}{\bibfnamefont{S.}~\bibnamefont{Havlin}},
  \bibinfo{journal}{Phys. Rev. Lett.} \textbf{\bibinfo{volume}{90}},
  \bibinfo{pages}{058701} (\bibinfo{year}{2003}).

\bibitem[{\citenamefont{J.~Leskovec and Faloutsos}()}]{leskovec06}
\bibinfo{author}{\bibfnamefont{J.~K.} \bibnamefont{J.~Leskovec}}
  \bibnamefont{and}
  \bibinfo{author}{\bibfnamefont{C.}~\bibnamefont{Faloutsos}},
  \bibinfo{note}{http://arxiv.org/physics/0603229}.

\bibitem[{\citenamefont{Newman}(2002)}]{n02a}
\bibinfo{author}{\bibfnamefont{M.~E.~J.} \bibnamefont{Newman}},
  \bibinfo{journal}{Phys. Rev. Lett.} \textbf{\bibinfo{volume}{89}},
  \bibinfo{pages}{208701} (\bibinfo{year}{2002}).

\bibitem[{\citenamefont{Harris}(2002)}]{harris02}
\bibinfo{author}{\bibfnamefont{T.~E.} \bibnamefont{Harris}},
  \emph{\bibinfo{title}{The Theory of Branching Processes}}
  (\bibinfo{publisher}{Springer-Verlag, Berlin}, \bibinfo{year}{2002}).

\bibitem[{\citenamefont{Jagers}(1975)}]{jagers75}
\bibinfo{author}{\bibfnamefont{P.}~\bibnamefont{Jagers}},
  \emph{\bibinfo{title}{Branching processes with biological applications}}
  (\bibinfo{publisher}{Wiley, London}, \bibinfo{year}{1975}).

\bibitem[{\citenamefont{Mode and Sleeman}(2000)}]{mode00}
\bibinfo{author}{\bibfnamefont{C.~J.} \bibnamefont{Mode}} \bibnamefont{and}
  \bibinfo{author}{\bibfnamefont{C.~K.} \bibnamefont{Sleeman}},
  \emph{\bibinfo{title}{Stochastic processes in epidemiology}}
  (\bibinfo{publisher}{World Scientific, Singapore}, \bibinfo{year}{2000}).

\bibitem[{\citenamefont{Godsil and Royle}(2001)}]{godsil01}
\bibinfo{author}{\bibfnamefont{C.}~\bibnamefont{Godsil}} \bibnamefont{and}
  \bibinfo{author}{\bibfnamefont{G.}~\bibnamefont{Royle}},
  \emph{\bibinfo{title}{Algebraic graph theory}} (\bibinfo{publisher}{Springer,
  New York}, \bibinfo{year}{2001}).

\bibitem[{\citenamefont{Lipsitch and {\it et al}}(2003)}]{lipsitch03}
\bibinfo{author}{\bibfnamefont{M.}~\bibnamefont{Lipsitch}} \bibnamefont{and}
  \bibinfo{author}{\bibnamefont{{\it et al}}}, \bibinfo{journal}{Science}
  \textbf{\bibinfo{volume}{300}}, \bibinfo{pages}{1966} (\bibinfo{year}{2003}).

\bibitem[{\citenamefont{Fraser et~al.}(2004)\citenamefont{Fraser, Riley,
  Anderson, and Ferguson}}]{fraser04}
\bibinfo{author}{\bibfnamefont{C.}~\bibnamefont{Fraser}},
  \bibinfo{author}{\bibfnamefont{S.}~\bibnamefont{Riley}},
  \bibinfo{author}{\bibfnamefont{R.}~\bibnamefont{Anderson}}, \bibnamefont{and}
  \bibinfo{author}{\bibfnamefont{N.~M.} \bibnamefont{Ferguson}},
  \bibinfo{journal}{Proc. Natl. Acad. Sci. USA} \textbf{\bibinfo{volume}{101}},
  \bibinfo{pages}{6146} (\bibinfo{year}{2004}).

\bibitem[{\citenamefont{Gantmacher}(1990)}]{gantmacher90}
\bibinfo{author}{\bibfnamefont{F.~R.} \bibnamefont{Gantmacher}},
  \emph{\bibinfo{title}{Mathrix Theory}} (\bibinfo{publisher}{AMS},
  \bibinfo{year}{1990}), \bibinfo{note}{vol. I}.

\bibitem[{\citenamefont{Galvani and May}(2005)}]{galvani05}
\bibinfo{author}{\bibfnamefont{A.~P.} \bibnamefont{Galvani}} \bibnamefont{and}
  \bibinfo{author}{\bibfnamefont{R.~M.} \bibnamefont{May}},
  \bibinfo{journal}{Nature} \textbf{\bibinfo{volume}{438}},
  \bibinfo{pages}{293} (\bibinfo{year}{2005}).

\bibitem[{\citenamefont{Vazquez}(2003)}]{vazquez03a}
\bibinfo{author}{\bibfnamefont{A.}~\bibnamefont{Vazquez}},
  \bibinfo{journal}{Phys. Rev. E} \textbf{\bibinfo{volume}{67}},
  \bibinfo{pages}{056104} (\bibinfo{year}{2003}).

\end{thebibliography}

\end{document}